\documentclass[aps,prl,preprint,twoside,floatfix]{revtex4}
\usepackage[dvips]{color}
\usepackage{graphicx, subfigure}
\usepackage{newlfont}

\begin{document}
\title{Reduction of anomalous heating in an $\emph{in-situ}$-cleaned ion trap  }
\author{D. A. Hite}
\author{Y. Colombe}
\author{A. C. Wilson}
\author{K. R. Brown}
\altaffiliation{Present Address: Georgia Tech Research Institute, Atlanta, Georgia  30318}
\author{U. Warring}
\author{R. J\"ordens}
\author{J. D. Jost}
\author{D. P. Pappas}
\author{D. Leibfried}
\author{D. J. Wineland}
\affiliation{National Institute of Standards and Technology, 325 Broadway, Boulder, Colorado  80305}

\begin{abstract}
Anomalous heating of trapped atomic ions is a major obstacle to their use as quantum bits in a scalable quantum computer.  The physical origin of this heating is not fully understood, but experimental evidence suggests that it is caused by electric-field noise emanating from the surface of the trap electrodes. In this study, we have investigated the role that adsorbates on the electrodes play by identifying contaminant overlayers, developing an \emph{in situ} argon-ion beam cleaning procedure, and measuring ion heating rates before and after cleaning the trap electrodes' surfaces.  We find a reduction of two orders of magnitude in heating rate after cleaning.
\end{abstract}
\maketitle

Trapped atomic ions can potentially be employed as quantum bits (qubits) in a scalable quantum computer, where multi-qubit logic gates require precise control of the ions' collective motion.  These operations incur errors caused by heating of the ions' motion from electric field noise of unknown origin. This noise has inhibited progress in scalability, miniaturization, and logic gate fidelity.  Research groups have addressed this problem by trying different electrode materials and processing techniques, but there can be wide variations in the observed heating.  Operation at low temperature can substantially reduce the heating [1, 2], but it is important to identify and eliminate the root causes.  The observed variations suggest that electrode surface contaminants may play a role.   Recently, application of a pulsed laser beam resulted in a reduction in heating rate by approximately a factor of two [3].  Here, we report a reduction in ion heating by two orders of magnitude, in a room-temperature surface-electrode ion trap [4] that has been cleaned \emph{in situ} by argon ion beam bombardment.   This suggests that anomalous heating can be significantly reduced or perhaps eliminated, without the need for, or in combination with, cryogenic cooling.

Ion heating is caused by electric-field noise at the location of the ion whose spectrum overlaps the frequency of the ions' motional modes. The physical origin of this noise has been debated for more than a decade.  Johnson noise is one source, but in many experiments, its contribution is estimated to be orders of magnitude smaller than the observed heating.   If the noise is caused by independently fluctuating potential patches on the electrodes that are small compared to the ion-electrode distance \emph{d}, the noise spectral density (proportional to the ion heating rate) is proportional to \emph{d}$^{-4}$ [5].   These potential fluctuations may be due to adsorbate-dipole fluctuations [6, 7], or to adatom-diffusion-induced work-function fluctuations on the electrode surface [8, 9].  Therefore, we have focused on removing contamination from the surface.

Cleaning of electrode surfaces was accomplished \emph{in situ} by Ar$^+$-ion bombardment.  Auger electron spectroscopy (AES) was performed in a separate surface analysis system on several duplicates of the ion-trap electrodes to determine their degree of cleanliness.   As seen in Fig.1-(a, top), after exposure to air and vacuum baking at 475 K (as used to prepare the ion-trap system), the electrode surfaces are covered with  2 - 3 monolayers of  oxygen-free carbon contamination.  After ion-beam cleaning (Fig. 1(a, bottom)), the absence of AES peaks not associated with Au indicates a clean surface.

To determine the electric-field noise, a $^9$Be$^+$ ion was trapped 40 $\mu$m above the trap electrodes, composed of 10-$\mu$m thick Au, electroplated onto a crystalline-quartz substrate.  After the ion is laser-cooled to the motional ground state, heating rate measurements [5] are made on a motional mode parallel to the trap surface, with a nominal frequency of $\omega$/2$\pi$ = 3.6 MHz.  The electric-field noise spectral density $S_E (\omega)$ and the heating rate in terms of rate of increase in motional quanta $\dot{\overline{n}}$ are related by [5]:
\begin{equation}
S_E(\omega) = {4m \hbar \omega \over q^2} \dot{\overline{n}},
\label{eq:spectral noise}
\end{equation}
where \emph{q} is the charge of the ion, \emph{m} is its mass, and $\hbar$ is Planck's constant divided by 2$\pi$.  After five cleaning attempts with increasing dosage, a subsequent cleaning, with 2 kV and 420 $\pm$  220 C/m$^2$ over 45 minutes at 5 $\times$  10$^{-3}$ Pa Ar pressure, reduced the heating rate from 7022 $\pm$  145 quanta/s to 58 $\pm$  2 quanta/s (Fig. 1 (b)).  The inferred $S_E$ (3.2 $\times$  10$^{-13}$ V$^2$/m$^2$/Hz) is comparable to the lowest values observed in cryogenic ion traps (Fig. 1(b)).

The dependence of $\dot{\overline{n}}$ on frequency can possibly give insight into the physical mechanisms responsible for the noise.  In many experiments, $\dot{\overline{n}}$ is seen to follow a 1/${\omega}^{\alpha}$ dependence, where values of $\alpha$ tend to group around 2.  Therefore, in Fig. 1(b) we plot  $\omega$$\cdot$$S_E(\omega)$ to approximately compensate the frequency dependence.  In this work, we observed a power-law dependence with $\alpha$  = 2.53 $\pm$  0.07 before cleaning and $\alpha$ = 2.57$\pm$   0.04 after cleaning, for $\omega$/2$\pi$  between 1.7 and 4.7 MHz, consistent with the surface-diffusion-noise model [8, 9] or certain parameter ranges of other models [6, 7].  This unchanged dependence may indicate that the noise is dominated by the same source before and after the cleaning, albeit significantly reduced. We conservatively estimate the Johnson-noise contribution for the resistances, filters, and geometry used in our trap to be $\sim$ 1 quantum/s, an additional 1.5 orders of magnitude below our post-cleaning result (Fig. 1 (b)).

This work was supported by IARPA, NSA, ONR, and the NIST Quantum Information Program.  We thank J. J. Bollinger and B. C. Sawyer for their suggestions on the manuscript.  This article is a contribution of the U.S. Government, and is not subject to U.S. copyright.

\pagebreak

\begin{figure}

           \includegraphics[width=.90\textwidth]{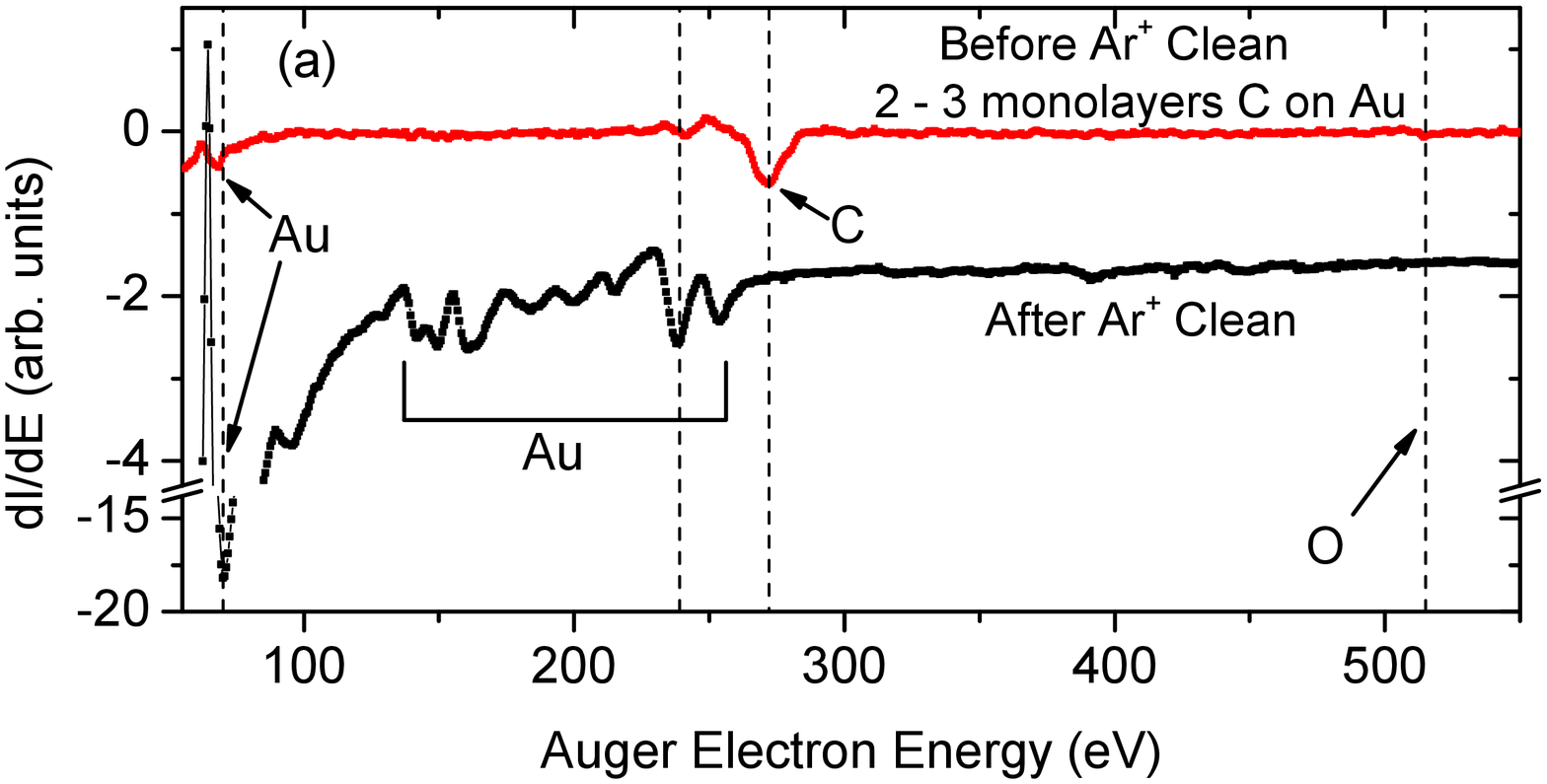}

          \includegraphics[width=1\textwidth]{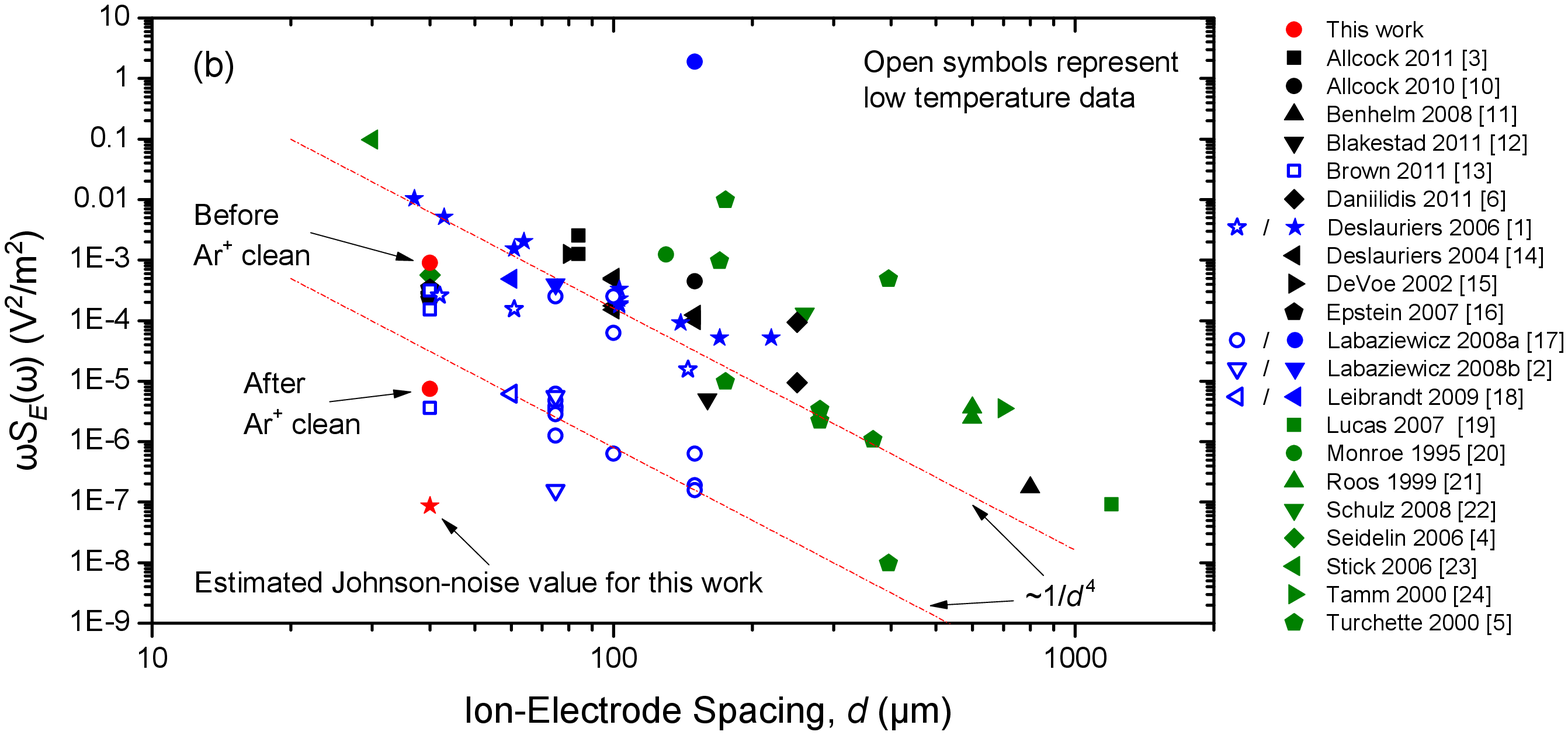}

    \caption{(a) Auger electron spectra of electrode surfaces.  The vertical axis displays the differential Auger electron intensity, dI/dE.  Before Ar$^+$-beam cleaning, carbon is observed as the only significant contaminant, probably resulting from oxygen-free hydrocarbon contamination.  After-Ar$^+$-Clean spectrum is offset vertically for clarity.  (b)  $\omega$$\cdot$$S_E(\omega)$ plotted versus ion-electrode distance, \emph{d}.   Our data (red filled circles) indicate a reduction of anomalous heating by two orders of magnitude after Ar$^+$-beam cleaning.  Dotted lines indicate a 1/\emph{d}$^4$ slope (vertical position not relevant).}
\end{figure}

\end{document}